\documentclass[prd,onecolumn]{revtex4}
\usepackage{dcolumn}
\usepackage{multirow}
\usepackage{graphicx}
\usepackage{amssymb}
\usepackage{bm}
\usepackage{hyperref}
\usepackage{epstopdf}
\usepackage{color}
\usepackage{mathrsfs}
\usepackage{amsmath,amssymb,amsthm}
\begin{document}
\title{Reconstructing $f(R)$ gravity from viscous cosmology constrained by observations}
\author{Deng Wang}
\email{Cstar@mail.nankai.edu.cn}
\affiliation{Theoretical Physics Division, Chern Institute of Mathematics, Nankai University,
Tianjin 300071, China}
\author{Xin-he Meng}
\email{xhm@nankai.edu.cn}
\affiliation{{Department of Physics, Nankai University, Tianjin 300071, P.R.China}\\
{State Key Lab of Theoretical Physics,
Institute of Theoretical Physics, CAS, Beijing 100080, P.R.China}}
\begin{abstract}
In this paper, we reconstruct the $f(R)$ theory using the viscous $\omega$CDM scenario (V-$\omega$CDM) constrained by several differently astrophysical observations, including Type Ia supernovae (SNe Ia), observational Hubble parameter data (OHD), Wilkinson Microwave Anisotropy Probe 9-year data (WMAP-9), Planck  observations and the single data point from the newest event GW150914 (GW). We find that the joint constraints from SNe Ia+OHD+Planck+GW data-sets give out a tighter restriction for the parameters of the V-$\omega$CDM scenario than other constraints. Subsequently, we find that there exists a substantially high degeneracy among these reconstructed $f(R)$ theories in the past, and that they start exhibiting differently evolutional behaviors around the present epoch as well as much different in the distant future. Furthermore, we make a comparison between the reconstructed $f(R)$ theory with the V-$\omega$CDM scenario and that with the $\omega$CDM scenario constrained by using the SNe Ia+OHD+Planck+GW data-sets in the $f(R)-R$ plane, and find that the two scenarios share an extremely high degeneracy, which implies that the effect of bulk viscosity plays a negligible role in the V-$\omega$CDM model during the whole evolutionary processes of the universe. In addition, one can also find that the reconstructed $f(R)$ theories using the two models just deviate from the standard Einstein gravity a little.

\end{abstract}
\maketitle
\section{Introduction}
The accelerating cosmic expansion is the most elegant and breathtaking discovery during the past 18 years \cite{1,2}, indicating either that the universe is dominated by an evolutionary dark component, namely, the so-called dark energy with some bizarre physical features, or that the general theory of relativity (GR) breaks down on the cosmological scales. To understand the profound implications of the nature and cosmological origin of accelerated expansion, a great deal of aspirant efforts have been made in recent years. Experimentally, more and more cosmological probes are proposed, including Type Ia supernovae (SNe Ia), baryonic acoustic oscillations (BAO), cosmic microwave background (CMB) radiation, observational Hubble parameter data (OHD), the abundance of galaxy clusters (AGC), strong and weak gravitational lensing (SGL/WGL), etc. Theoretically, the succinctest candidate is the so-called $\Lambda$CDM scenario, which has been verified to be substantially successful in describing many aspects of the observed universe. For instance, the large scale structure (LSS) of matter distribution at the linear level, the spectrum of anisotropies of the CMB radiation, and the expansion phenomenon are very well depicted by the standard cosmological scenario. The newest results of Planck indicate there still exist some anomalies which are inconsistent with the predictions of the $\Lambda$CDM scenario \cite{p}, containing the anomalies of the observed Hubble parameter $H(z)$ and the amplitude of fluctuation spectrum. Besides these anomalies,
this scenario has confronted two fatal problems, i.e., the `` fine-tuning '' problem and the `` coincidence '' problem \cite{3}. The former implies that the theoretical value for the vacuum density are much greater than its observed value, namely, the famous 120-orders-of-magnitude discrepancy that makes the vacuum explanation puzzling, while the latter indicates that why dark matter and dark energy are at the same order today since their energy densities are so distinct during the evolutional process of the universe. Additionally, E. Witten has pointed out that a positive cosmological constant is incompatible with the perturbed string theory \cite{4}. Hence, the actual nature of dark energy may not be the the cosmological constant $\Lambda$. Based on this concern, floods of candidates have been proposed by cosmologists, in order to alleviate or even solve the aforementioned problems, containing quintessence \cite{5,6,7,8,9,10,11,12}, phantom \cite{13}, dark fluid \cite{14,15,16,17,18,19}, decaying vacuum \cite{20}, braneworld model \cite{21,22,23}, Chaplygin gas \cite{24}, f(R) gravity \cite{25,26,27,28}, scalar-tensor gravity \cite{29,30,31,32,33}, Einstein-Aether gravity \cite{34,35}, Chern-Simons gravity \cite{36}, etc.

As is well known, when GR breaks down, one would like to consider the modified theories of gravity (MOG) corresponding to modifications to the Einstein-Hilbert action. According to this point of view, one may take into account the high-order derivative theories of gravity, where the high-order curvature invariants (such as $R^2$, $R^{\mu\nu}R_{\mu\nu}$, $R^{\mu\nu\alpha\beta}R_{\mu\nu\alpha\beta}$, or $R\square^nR$), and nonminimally coupled terms between spacetime geometry and scalar fields (such as $\phi^2R$) usually emerge as quantum corrections in string theory or low energy effective action of quantum gravity \cite{37}. Naturally, the simplest case following this logical line is the so-called f(R) gravity where the modification is just a function of the Ricci scalar. In this case, the modified Friedmann Equations will be obtained by varying a generalized Lagrangian which is the aforementioned function of the Ricci scalar. It is very clear that the usual GR can be recovered in the limit $f(R)=R$, at the same time, completely different results could be acquired through other choices of f(R). Generally speaking, a mature f(R) theory of gravity should be responsible for the inflationary behavior in the very early universe, be suitable to present the late-time cosmic acceleration phenomenon, and satisfy the stability conditions. Subsequently, a question occurs, namely, what are the viable conditions for a thoughtful f(R) theory ? Here we would like to answer this question in the metric formalism as follows :

$\star$ $f'(R)>0$ for $R\geqslant R_0>0$, where $f'(R)\equiv\partial f/\partial R$ and $R_0$ denotes the present-day value of the Ricci scalar. It is noteworthy that, if the final attractor is a de Sitter point with the Ricci scalar $R_1$ ($R_1>0$), this condition needs to hold for $R\geqslant R_1$.

$\star$ $f(R)\rightarrow R-2\Lambda$ for $R\gg R_0$. This condition should be consistent with the so-called local gravity constraints \cite{38,39} and for the existence of the matter-dominated stage \cite{40}.

$\star$ $f''(R)>0$ for $R>R_0$, where $f''(R)\equiv\partial^2f/\partial R^2$. The condition should be consistent with local gravity constraints, for the existence of the matter-dominated stage, and for the stability of cosmological perturbations.

$\star$ $1>\frac{Rf''(R)}{f'(R)}\mid_{r=-2}>0$ at $r=-\frac{Rf'(R)}{f}=-2$. This condition is required for the stability of the de Sitter point in the late-time universe.

For instance, in the literature, two famous f(R) models, namely, the the Starobinsky model and Hu-Sawicki model, have been proposed, and they satisfy the above-mentioned four conditions substantially well.

In general, once the Friedmann-Lema\^{i}tre-Robertson-Walker (FLRW) spacetime geometry is assumed, there are usually two approaches to derive the field equations governing the dynamics of the universe, i.e., the metric formalism and the Palatini formalism. In the former case, one obtains the field equations by varying the action with respect to the metric only, while in the latter case, one needs to vary the action with respect to both the metric and the connections components. The two methods will give out the same result only in the case $f(R)=R$, while derive, usually, completely different field equations for other choices of the Lagrangian. The Palatini formalism has a more significant advantage than the metric formalism, from which the derived dynamical equations is just of second order. However, in the case of the metric formalism, the dynamical equations will be fourth order for the scalar factor $a(t)$, and can not be solved exactly in most cases. Therefore, it is natural that one may attempt to solve the field equations numerically, but, unfortunately, there exists substantially large uncertainties on the parameters determining the boundary conditions. Recently, in the paper \cite{41}, S. Capozziello et al. have depicted the possibility to obtain a numerical solution in the metric formalism. They consider the dynamical equations as a linear third-order differential equation for $f[R(z)]$ as the function of the redshift $z$. In addition, using this method, it is easy to set the boundary conditions for $f[R(z)]$ on the basis of very physical considerations and get rid of the influence from the measurement errors. Objectively speaking, first of all, for such a method one must determine the Hubble parameter $H(z)$ from the corresponding cosmological models constrained by astronomical observations. In the second place, one can directly reconstruct the concrete $f(R)$ theory of gravity from the Hubble parameter $H(z)$, since $R(z)$ can be expressed as the function of $H(z)$ and its derivative. In total, one can determine what is the corresponding $f(R)$ theory of gravity when $H(z)$ is determined for a concrete dark energy scenarios.

In the present situation, we would like to use the viscous cosmological scenarios to reconstruct the $f(R)$ theory of gravity, and explore the related properties of the reconstructed $f(R)$ theory. As is well known, the dynamics of a realistic physical system can not be governed only by a perfect fluid, and all the observations indicate that the universe media is not a perfect fluid at the present epoch. Hence, one should take into account the dissipative effect of the cosmological fluid for a more realistic model. About the cosmological models including the bulk viscosity, we refer the readers to the paper \cite{19}, which contain systematic and detailed description about the features of viscous cosmology. As a consequence, we are full of interest to reconstruct $f(R)$ theory of gravity from the bulk viscosity cosmology, which is not studied wholly by other authors in the previous literature. To be more precise, in this paper, we will reconstruct the $f(R)$ theory from the viscous $\omega$CDM (V-$\omega$CDM) scenario (i.e., viscous quiessence scenario).

This paper is organized as follows. In Section 2, we would like to review the V-$\omega$CDM cosmological model briefly. In Section 3, we will place constraints on the V-$\omega$CDM cosmological model by adopting the SNe Ia, OHD, CMB and the single data point from the newest event GW150914 \cite{42}. Reconstructing the $f(R)$ theory of gravity from the V-$\omega$CDM cosmological model will be explored in Section 4. In the final section, the discussions and conclusions are presented (We take units $8\pi G=c=\hbar=1$ throughout the context).

\section{Review on V-$\omega$CDM cosmological model}
Consider the concrete case where the spacetime is undergoing the phase of thermal equilibrium and without the shear viscosity. Furthermore, using the spatially flat FLRW metric, the stress-energy tensor can be expressed as
\begin{equation}
T_{\mu\nu}=\rho U_{\mu}U_{\nu}+(p-\theta\zeta)h_{\mu\nu}, \label{1}
\end{equation}
where $\rho$ denotes the mass-energy density, p the isotropic pressure, $U_{\mu}=(1, 0, 0, 0)$ the four-velocity of the cosmic fluid in comoving coordinates, $\theta=3H$ the expansion scalar, $\zeta$ the bulk viscosity and $h_{\mu\nu}$ the projection tensor. By defining the effective pressure $\tilde{p}=p-\theta\zeta$ and starting from the Einstein field equations, one can easily obtain the Friedmann equations as follows :
\begin{equation}
\frac{\dot{a}^2}{a^2}=\frac{\rho}{3},\label{2}
\end{equation}
\begin{equation}
\frac{\ddot{a}}{a}=-\frac{\rho}{6}(\rho+3\tilde{p}).\label{3}
\end{equation}
Note that we have set $8\pi G=1$. The energy conservation equation, namely, $T^{0\nu}_{\hspace{3mm};\nu} = 0$, can be written as
\begin{equation}
\dot{\rho}+3H(\rho+\tilde{p})=0,\label{4}
\end{equation}
where $H=\dot{a}/a$ denotes the Hubble expansion rate. Furthermore, if one takes into account the V-$\omega$CDM scenario and sets the bulk viscosity coefficient as a constant $\tau$ (namely, $\zeta=\tau H$), the effective pressure will be $\tilde{p}=\omega\rho-3\tau H^2$. As a consequence, the corresponding Friedmann equations, namely, Eq. (\ref{4}) can be rewritten as :
\begin{equation}
\dot{\rho}+3H[(1+\omega)\rho-3\tau H^2]=0.\label{5}
\end{equation}
Combining Eq. (\ref{1}) with Eq. (\ref{5}) and neglecting the radiation contribution, one can obtain the dimensionless Hubble parameter as follows

\begin{equation}
E(z)=[\Omega_{m0}(1+z)^3+\Omega_{de0}(1+z)^{3(1+\omega-\tau)}]^{1/2},   \label{6}
\end{equation}
where $\Omega_{m0}$ is the present value of the matter density parameter and $\Omega_{de0}$ the present value of the dark energy density parameter. It is easy to be checked that when the viscosity term vanishes ($\tau=0$), Eq. (\ref{6}) will reduce to the case of $\omega$CDM scenario. In the next section, we will constrain the V-$\omega$CDM scenario by using the SNe Ia, OHD, CMB and the single data point from the newest event GW150914.

\section{Methodology and constraint results}
\subsection{Type Ia Supernovae}
In modern cosmology, SNe Ia luminosities have been corrected and used as standard candles to probe the expansion history of the universe. As is mentioned above, this technique have been successfully applied into discovering the present accelerating phenomenon. With the gradually increasing number of high redshift SNe Ia and statistical methods are improved step by step, one can also give out a tighter constraint than previous results for a given dark energy model in the future. The Union 2.1 data-sets are taken for numerical fitting, consisting of 580 SNe Ia data points. To perform the standard $\chi^2$ statistics, the predicted distance modulus for a given set of model parameters $\theta$, can be defined as follows
\begin{equation}
\mu_{p}(z_i)=m-M=5\log_{10}D_L(z_i)+25, \label{7}
\end{equation}
where $m$ denotes the apparent magnitude, $M$ the absolute magnitude as well as $D_L(z_i)$ the luminosity distance at a given redshift $z_i$ in units of Mpc,
\begin{equation}
D_L(z_i)=(1+z_i)\int^{z_i}_0\frac{dz'}{E(z';\theta)}, \label{8}
\end{equation}
where $E(z';\theta)$ is the dimensionless Hubble parameter. Then one can directly express the $\chi^2$ for the SNe Ia observations as follows
\begin{equation}
\chi^2_S=\sum^{580}_{i=1}[\frac{\mu_{obs}(z_i)-\mu_{p}(z_i;\theta)}{\sigma_i}]^2,  \label{9}
\end{equation}
where $\mu_{obs}$ denotes the observed value, and $\sigma_i$ the corresponding 1¦Ò error of the observed distance modulus, respectively, at a given redshift $z_i$.
\subsection{Observational Hubble Parameter}
In the present situation, we would like to use the latest OHD data-sets to constraint the V-$\omega$CDM cosmological scenario.
The OHD data-sets have been used to constrain various kinds of cosmological scenarios, for instance, in our previous works \cite{43,44,45,46,47}, we have constrained a series of dark energy models including the Ricci dark energy model (RDE), Holographic dark energy model (HDE), time-dependent dark energy model (TDDE), Cardassian model as well as two parametric models for effective pressure. It is worth noticing that the OHD has a unique advantage, i.e., they are obtained from model-independent direct observations and avoid integrating over the redshift $z$ so as not to drop any useful information. So far, two methods in the literature have been developed to measure the OHD, namely, `` galaxy differential age '' and `` radial BAO size '' methods. More useful and detailed information can be found in paper \cite{48}. Then, the expected $\chi^2$ for the OHD can be defined in the following manner :
\begin{equation}
\chi^2_{H}=\sum^{36}_{i=1}[\frac{H_0E(z_i)-H_{obs}(z_i)}{\sigma_i}]^2 \label{10},
\end{equation}
where $H_0$ is the present value of the Hubble parameter and $H_{obs}(z_i)$ the observed value of the OHD at a given redshift $z_i$.

\subsection{Cosmic Microwave Background Anisotropy}
As is well known, another important and effective probe is the so-called CMB anisotropy. More recently, the WMAP collaboration released their final 9-year data-sets (WMAP-9) \cite{49}. Meanwhile, the Planck Collaboration also released their data-sets (Planck) \cite{50}, and they claimed there exists a subtle tension between WMAP-9 and Planck data-sets. Hence, in the present situation, we would like to consider the two data-sets separately. In addition, for simplicity, we will adopt the shift parameter $\mathcal{R}$ instead of the whole data from CMB to constrain the V-$\omega$CDM scenario, since using the full data of CMB to perform a global fitting will spend a large amount of power and computation time. It is also discussed in papers \cite{51} that the shift parameter $\mathcal{R}$ has contained the main information of the whole CMB data. The shift parameter $\mathcal{R}$ can be defined as
\begin{equation}
\mathcal{R}=\sqrt{\Omega_{m0}}\int^{z_C}_0\frac{dz'}{E(z')}, \label{11}
\end{equation}
where $z_C$ denotes the redshift of recombination. It has been determined that $\mathcal{R}=1.7302\pm0.0169$ and $z_C=1089.09$ for WMAP-9 \cite{49}, while $\mathcal{R}=1.7499\pm0.0088$ and $z_C=1090.41$ for Planck \cite{50}. The corresponding $\chi^2$ for the CMB observations can be defined as
\begin{equation}
\chi^2_{C}(\theta)=[\frac{\mathcal{R}_{obs}-\mathcal{R}(\theta)}{\sigma_{\mathcal{R}}}]^2, \label{12}
\end{equation}
where $\mathcal{R}_{obs}$ and $\sigma_{\mathcal{R}}$ correspond to the value of the shift parameter and the value of $1\sigma$ error extracted from the aforementioned two data-sets, respectively.

\subsection{Gravitational Wave}
The LIGO detection of the gravitational wave transient GW150914, from the inspiral and merger of two black holes with masses $\gtrsim 30$ $M_{\odot}$, not only indicates that the existence of supermassive black holes but also successfully ushers in a new era of multi-messenger astronomy. In this situation, we will make the first interesting try, i.e., transforming the single data point to an available SNe Ia data point and calculating out the corresponding distance modulus $38.0639^{+0.7155}_{-1.2553}$. Though the quality the single data point at present is not very well, we believe strongly the forthcoming gravitational-wave observations can provide better data-sets and open a window for new physics. For simplicity, we would like to denote the corresponding statistical contribution $\chi^2$ for the GW150914 as $\chi^2_G$.

\begin{figure}
\centering
\includegraphics[scale=0.5]{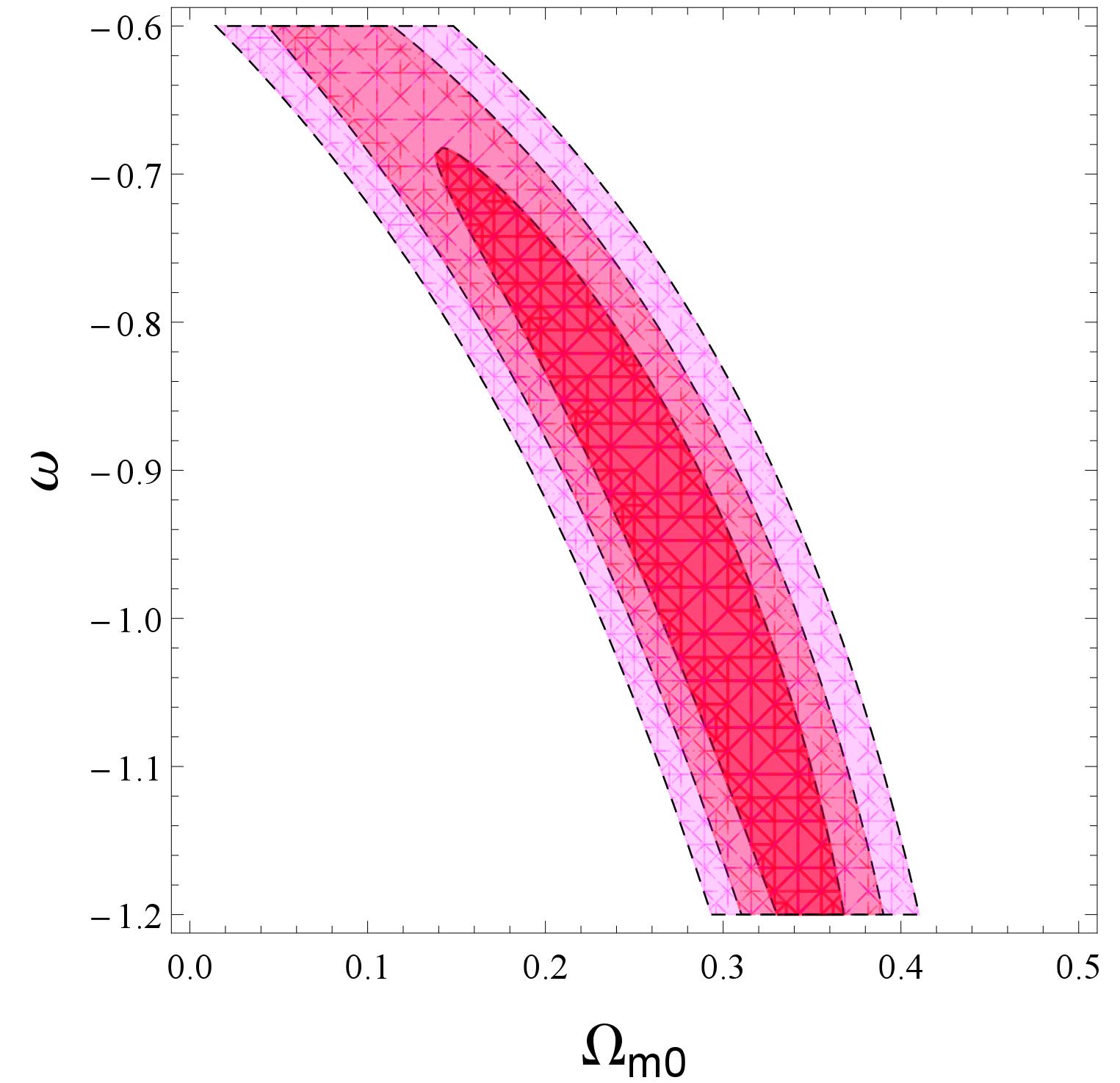}
\caption{The $1\sigma$, $2\sigma$ and $3\sigma$ confidence intervals for the parameter pair ($\Omega_{m0},\omega$) of V-$\omega$CDM model constrained only
by SNe Ia data-sets. Here we have chosen the best fitting value $\tau=0.0446$ of the bulk viscosity coefficient.}\label{fa1}
\end{figure}
\begin{figure}
\centering
\includegraphics[scale=0.5]{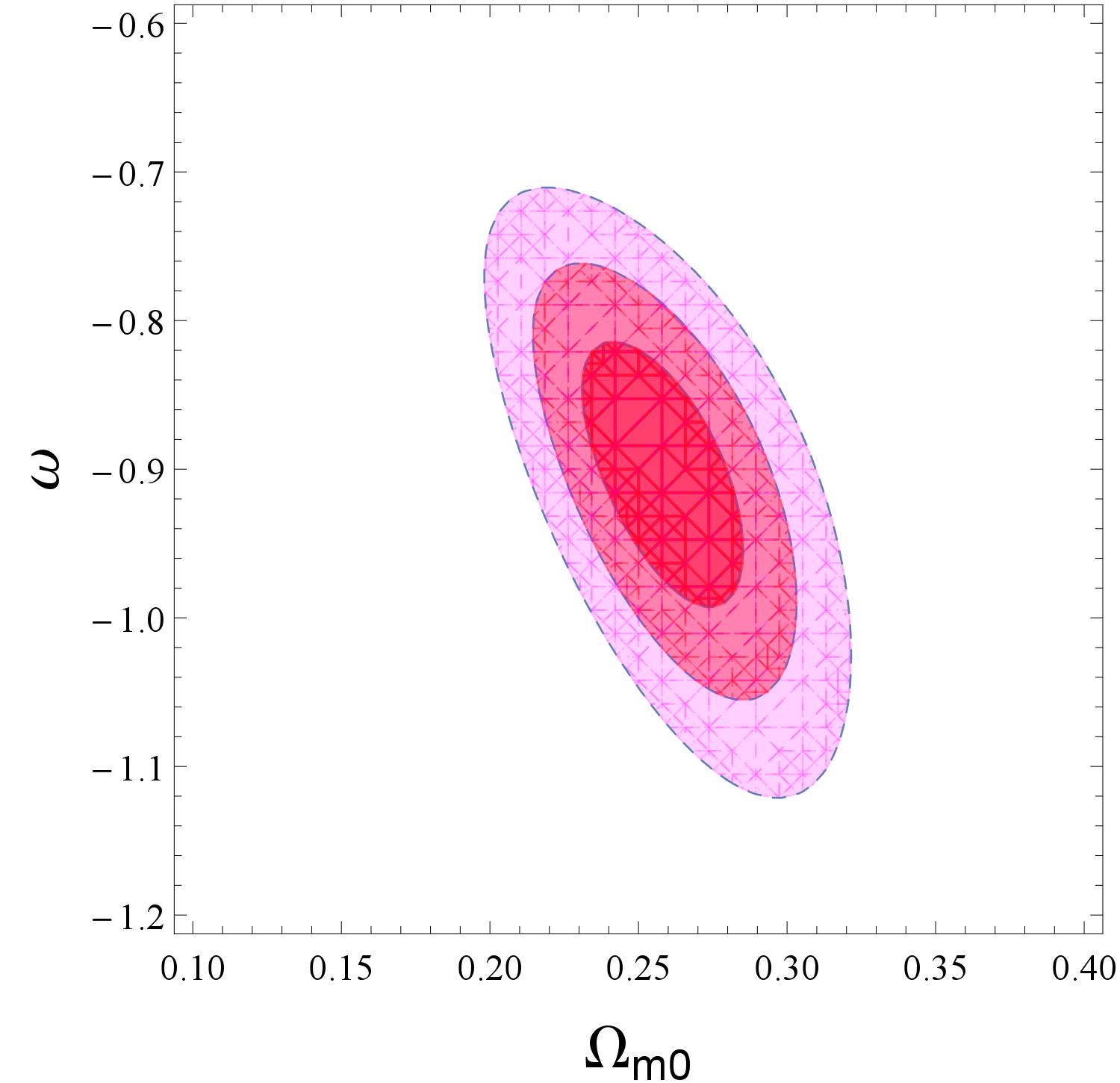}
\caption{The $1\sigma$, $2\sigma$ and $3\sigma$ confidence intervals for the parameter pair ($\Omega_{m0},\omega$) of V-$\omega$CDM model constrained
by SNe Ia+OHD data-sets. Here we have chosen the best fitting value $\tau=0.0517$ of the bulk viscosity coefficient.}\label{fb1}
\end{figure}
\begin{figure}
\centering
\includegraphics[scale=0.5]{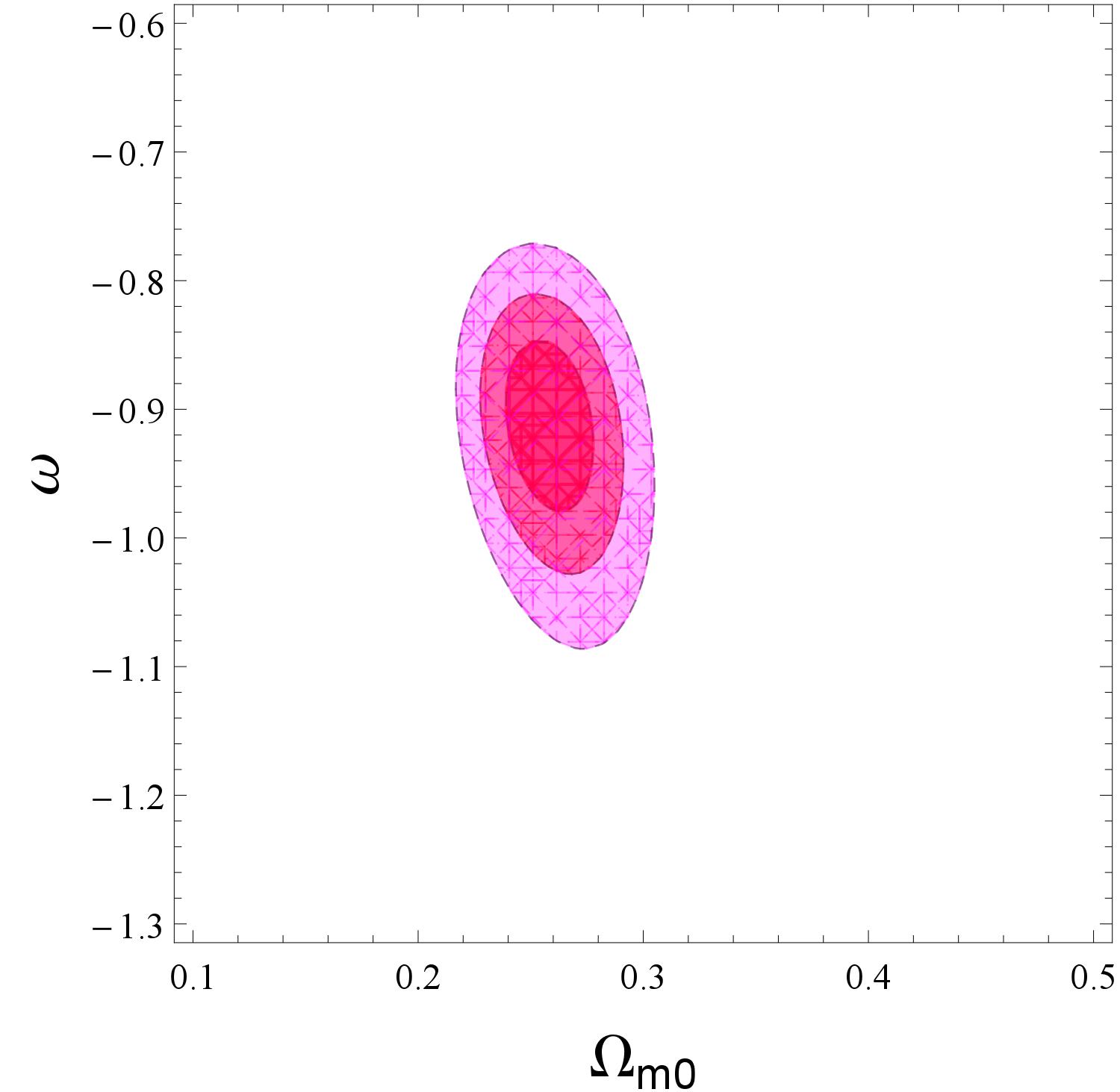}
\caption{The $1\sigma$, $2\sigma$ and $3\sigma$ confidence intervals for the parameter pair ($\Omega_{m0},\omega$) of V-$\omega$CDM model constrained
by SNe Ia+OHD+WMAP-9+GW data-sets. Here we have chosen the best fitting value $\tau=0.0626$ of the bulk viscosity coefficient.}\label{fc1}
\end{figure}
\begin{figure}
\centering
\includegraphics[scale=0.5]{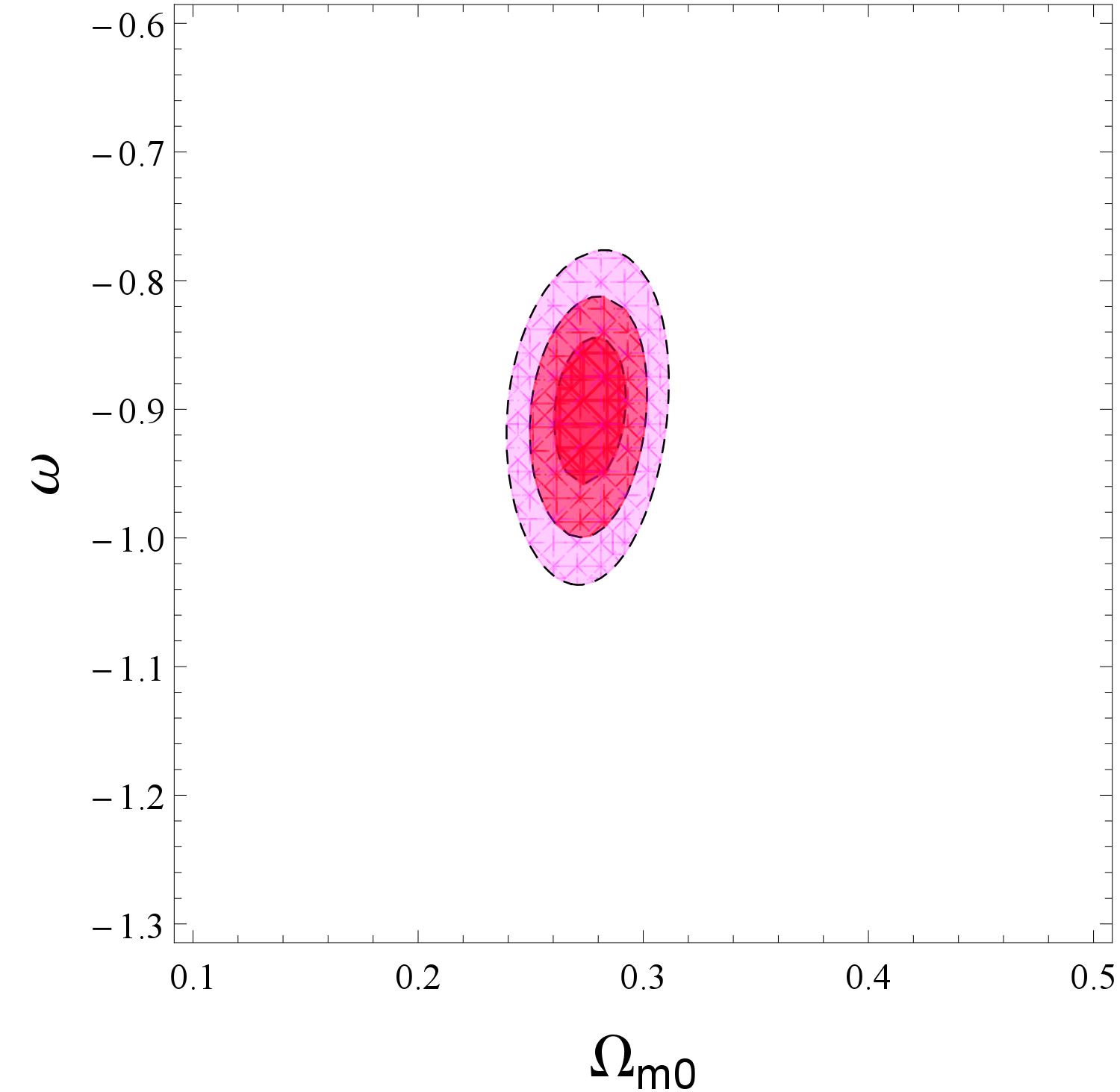}
\caption{The $1\sigma$, $2\sigma$ and $3\sigma$ confidence intervals for the parameter pair ($\Omega_{m0},\omega$) of V-$\omega$CDM model constrained
by SNe Ia+OHD+Planck+GW data-sets. Here we have chosen the best fitting value $\tau=0.1401$ of the bulk viscosity coefficient.}\label{fd1}
\end{figure}

Subsequently, the corresponding $\chi^2$ of the joint constraints from the SNe Ia, OHD, CMB as well as the GW150914 data-sets can be expressed as
\begin{equation}
\tilde{\chi}^2={\chi}^2_{S}+{\chi}^2_{H}+\chi^2_{C}++\chi^2_G \label{15}.
\end{equation}
Both the best fitting values of the model parameters and the minimal values $\chi^2_{min}$ of the derived $\tilde{\chi}^2$ from different joint constraints, for the V-$\omega$CDM model and $\omega$CDM model and are listed in Tables. \ref{tab1} and \ref{tab2}, respectively. Additionally, the likelihood distributions of different constraints for the V-$\omega$CDM model are depicted in Figures. \ref{fa1}-\ref{fd1}. It is not difficult to discover that the joint constraint from SNe Ia+OHD+Planck+GW data-sets gives out a tighter restriction for the V-$\omega$CDM model than other constraints. In the next section, we would like to use the best fitting values of the V-$\omega$CDM model to reconstruct the corresponding $f(R)$ theory, and investigate the related characteristics and evolutional tendency.
\begin{table}[h!]
\begin{tabular}{|c|c|c|c|c|cc}
\hline
                      &SNe Ia           &SNe Ia+OHD       &SNe Ia+OHD+WMAP-9+GW           &SNe Ia+OHD+Planck+GW\\
\hline
$\chi^2_{min}$        & $562.224$       &$579.451$      &$579.873$        &$582.169$\\
$\Omega_{m0}$         & $0.2810$      &$0.2578$      & $0.2647$     & $0.2815$\\
$\omega$            & $-0.9646$       &$-0.9011$      & $-0.9132$     & $-0.8908$ \\
$\tau$            &$0.0446$        &$0.0517$      &$0.0626$      &$0.1401$\\
\hline
\end{tabular}
\caption{The best fitting values of the model parameters ($\Omega_{m0}$, $\omega$, $\tau$) in the V-$\omega$CDM model by adopting the joint constraints from the SNe Ia, OHD, WMAP-9 (or Planck) and the gravitational-wave data-sets.}
\label{tab1}
\end{table}

\begin{table}[h!]
\begin{tabular}{|c|c|c|c|c|cc}
\hline
                      &SNe Ia           &SNe Ia+OHD       &SNe Ia+OHD+WMAP-9+GW           &SNe Ia+OHD+Planck+GW\\
\hline
$\chi^2_{min}$        & $562.227$       &$579.451$      &$579.873$        &$582.169$\\
$\Omega_{m0}$         & $0.2848$      &$0.2577$      & $0.2647$     & $0.2815$\\
$\omega$            & $-1.0194$       &$-0.9530$      & $-0.9757$     & $-0.1.0308$ \\
\hline
\end{tabular}
\caption{The best fitting values of the model parameters ($\Omega_{m0}$, $\omega$, $\tau$) in the $\omega$CDM models by adopting the joint constraints from the SNe Ia, OHD, WMAP-9 (or Planck) and the gravitational-wave data-sets.}
\label{tab2}
\end{table}
\section{Reconstructing $f(R)$ gravity from V-$\omega$CDM scenario}
Consider the modified theories of gravity, one can incorporate terms into the effective Lagrangian of the gravitational field, such as $R^2$, $R^{\mu\nu}R_{\mu\nu}$, $R^{\mu\nu\alpha\beta}R_{\mu\nu\alpha\beta}$, or $R\square^nR$, when the quantum corrections are taken into account. In the case of $f(R)$ theory, the modification is just a function of Ricci scalar curvature, and the corresponding action can be expressed as
\begin{equation}
S=\int d^4x\sqrt{-g}[f(R)+\mathcal{L}_m], \label{16}
\end{equation}
where $f(R)$ is a function of $R$, $\mathcal{L}_m$ the standard matter Lagrangian and $g$ the trace of the corresponding metric. It is noteworthy that the ansatz $f(R)=R+2\Lambda$ reproduces the standard GR with a cosmological constant. By varying with respect to the metric components, the modified Einstein field equations can be expressed as follows \cite{52}:
\begin{equation}
G_{\mu\nu}=R_{\mu\nu}-\frac{1}{2}Rg_{\mu\nu}=T^{(m)}_{\mu\nu}+T^{(cur)}_{\mu\nu}, \label{17}
\end{equation}
where $G_{\mu\nu}$ denotes the Einstein tensor and the stress energy tensor of the matter
\begin{equation}
T^{(m)}_{\mu\nu}=\frac{\tilde{T}^{(m)}_{\mu\nu}}{f'(R)} \label{18}
\end{equation}
and the stress energy tensor for the effective curvature fluid
\begin{equation}
T^{(cur)}_{\mu\nu}=\frac{1}{f'(R)}\{\frac{g_{\mu\nu}}{2}[f(R)-Rf'(R)]+f'(R)^{;\mu\nu}(g_{\mu\rho}g_{\nu\beta}-g_{\mu\nu}g_{\rho\beta})\}, \label{18}
\end{equation}
where $\tilde{T}^{(m)}_{\mu\nu}$ is the standard minimally coupled stress energy tensor. For the flat FLRW metric, one can obtain the modified Friedmann equations as follows \cite{52}:
\begin{equation}
H^2=\frac{1}{3}[\rho_{cur}+\frac{\rho_m}{f'(R)}], \label{19}
\end{equation}
\begin{equation}
H^2+2\frac{\ddot{a}}{a}=-(\rho_{cur}+\rho_m), \label{20}
\end{equation}
where $\rho_m$ and $p_m$ are the matter energy density and the corresponding pressure, respectively.  According to the paper \cite{41}, the same physical quantities for the effective curvature fluid are
\begin{equation}
\rho_{cur}=\frac{1}{f'(R)}\{\frac{1}{2}[f(R)-Rf'(R)]-3H\dot{R}f''(R)\} \label{21}
\end{equation}
and
\begin{equation}
p_{cur}=\frac{1}{f'(R)}\{2H\dot{R}f''(R)+\ddot{R}f''(R)+\dot{R}^2f'''(R)+\frac{1}{2}[Rf'(R)-f(R)]\}. \label{22}
\end{equation}
Using the Bianchi identity to Eq. (\ref{17}), one can easily obtain the conservation equation for the total energy density as follows :
\begin{equation}
\dot{\rho}_{tot}+3H(\rho_{tot}+p_{tot})=0. \label{23}
\end{equation}
It is worth noting that we consider the dust matter (i.e., $p_m=0$) only throughout the context, and do not take into account the interaction between the curvature fluid and the matter. Therefore, one can naturally assume the matter energy density is conserved in the process of reconstruction in order that
\begin{equation}
\rho_m=3H_0^2\Omega_{m0}(1+z)^3, \label{24}
\end{equation}
where $z=1/a-1$ and $a(t_0)=1$ ($t_0$ represents the present epoch). Furthermore, since Eqs. (\ref{20}), (\ref{21}) and (\ref{24}) are not mutually independent, the dynamics of the universe can be completely determined by Eqs. (\ref{20}) and (\ref{24}). Then, the system can be described conveniently by the following equation :
\begin{equation}
\dot{H}=-\frac{1}{2f'(R)}\{3H_0^2\Omega_{m0}(1+z)^3+\ddot{R}f''(R)+\dot{R}[\dot{R}f'''(R)-Hf''(R)]\}, \label{25}
\end{equation}
As is mentioned in the introduction, for the purpose to reconstruct, it is useful to change the differential variable from the cosmic time to the redshift $z$ depending on the following relation :
\begin{equation}
\frac{d}{dt}=-(1+z)H\frac{d}{dz}. \label{26}
\end{equation}
The Ricci scalar curvature in the flat universe can be written as
\begin{equation}
R=6[(1+z)H\frac{dH}{dz}-2H^2]. \label{27}
\end{equation}
Subsequently, combining the Eqs. (\ref{26}) and (\ref{27}), Eq. (\ref{25}) can be expressed as a third-order differential equation of $f(z)$ ($f[R(z)]=f(z)$) :
\begin{equation}
\mathcal{D}_3(z)\frac{d^3}{dz^3}+\mathcal{D}_2(z)\frac{d^2}{dz^2}+\mathcal{D}_1(z)\frac{d}{dz}=-3H_0^2\Omega_{m0}(1+z)^3 \label{28}
\end{equation}
with :
\begin{equation}
\mathcal{D}_1(z)=\dot{R}^2(\frac{dR}{dz})^{-4}[3(\frac{d^2R}{dz^2})^2(\frac{dR}{dz})^{-1}-\frac{d^3R}{dz^3}]+(\dot{R}H-\ddot{R})\frac{d^2R}{dz^2}(\frac{dR}{dz})^{-3}-2(1+z)H\frac{dH}{dz}(\frac{dR}{dz})^{-1}, \label{29}
\end{equation}
\begin{equation}
\mathcal{D}_2(z)=(\ddot{R}-\dot{R}H)(\frac{dR}{dz})^{-2}-3\dot{R}^2(\frac{dR}{dz})^{-4}\frac{d^2R}{dz^2}, \label{30}
\end{equation}
and
\begin{equation}
\mathcal{D}_3(z)=\dot{R}^2(\frac{dR}{dz})^{-3}. \label{31}
\end{equation}

\begin{figure}
\centering
\includegraphics[scale=0.5]{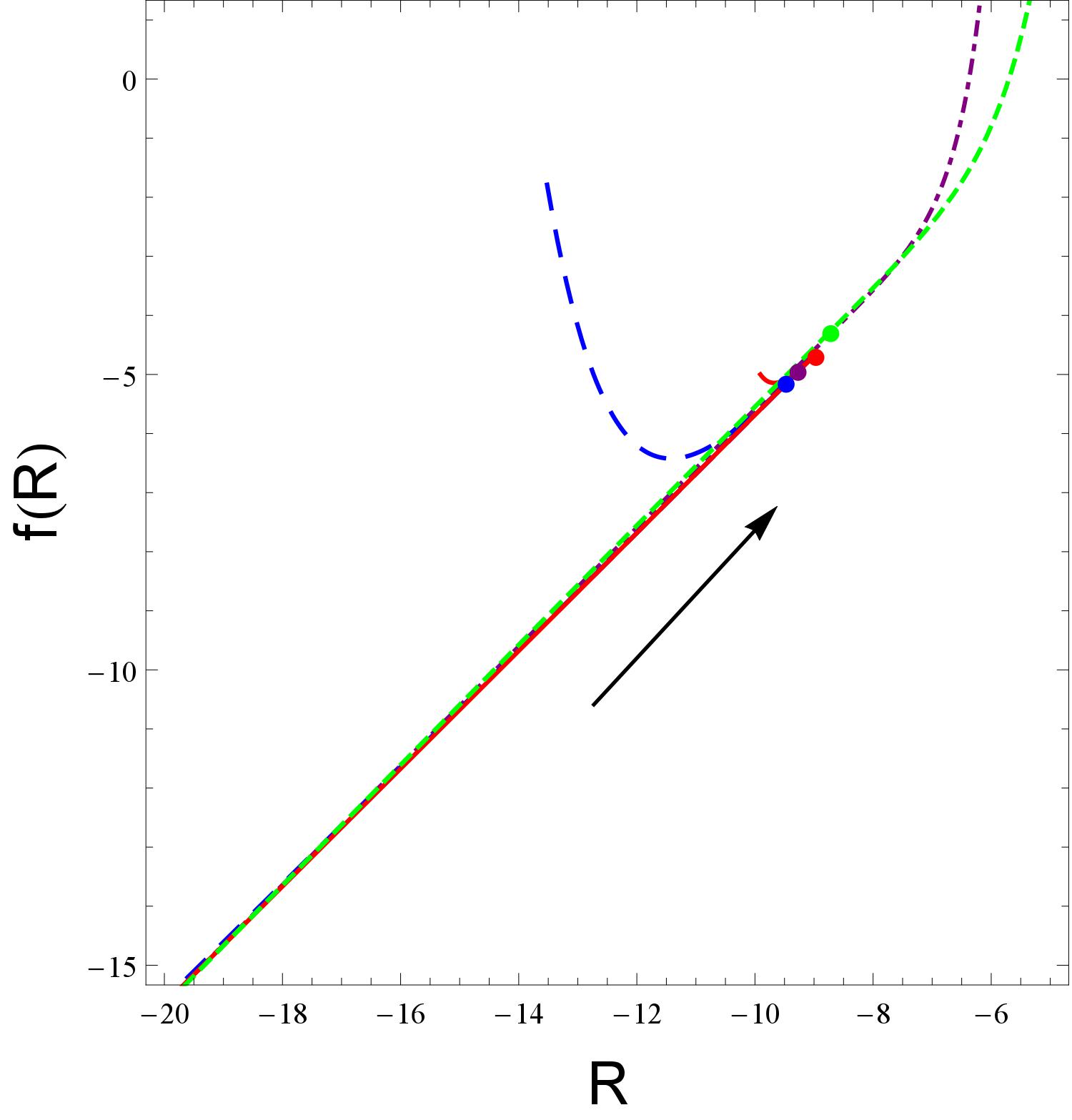}
\caption{Reconstructed $f(R)$ with V-$\omega$CDM scenario in the range $z\in(-1,1.5]$, where the red (solid) line represents the reconstructed $f(R)$ theory  from the SNe Ia data-sets, the green (short-dashed) line from SNe Ia+OHD, the purple (dash-dotted) line from SNe Ia+OHD+WMAP-9+GW as well as the blue (long-dashed) line from SNe Ia+OHD+Planck+GW. The black arrow indicates the evolutional direction of the V-$\omega$CDM scenario and the present epoch of reconstructed f(R) is shown as a dot.}\label{f1}
\end{figure}
\begin{figure}
\centering
\includegraphics[scale=0.5]{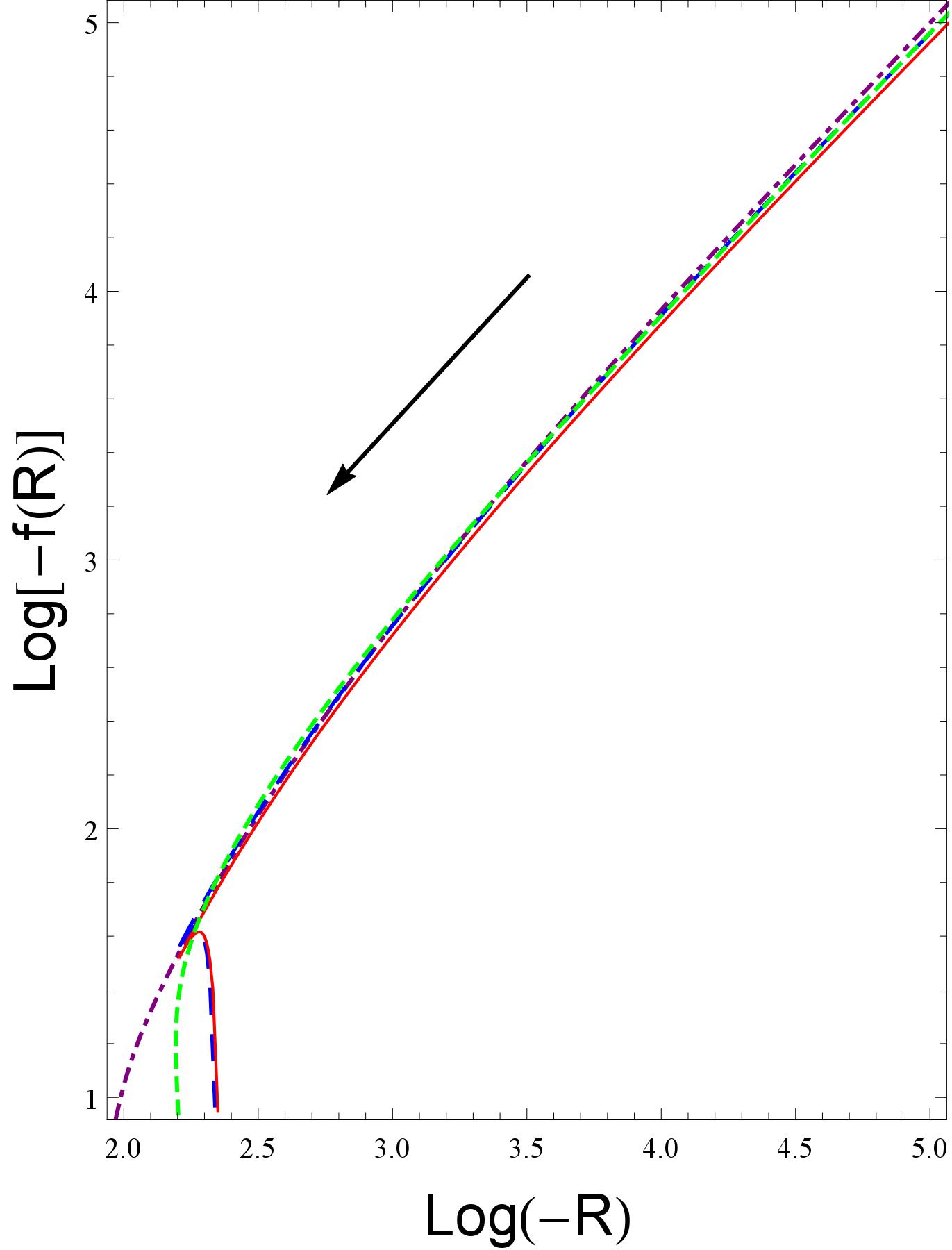}
\caption{Reconstructed $f(R)$ with V-$\omega$CDM scenario in the plane of $lf-lR$ (for the interval $z\in(-1,1.5]$), where the red (solid) line represents the reconstructed $f(R)$ theory  from the SNe Ia data-sets, the green (short-dashed) line from SNe Ia+OHD, the purple (dash-dotted) line from SNe Ia+OHD+WMAP-9+GW as well as the blue (long-dashed) line from SNe Ia+OHD+Planck+GW. The black arrow indicates the evolutional direction of the V-$\omega$CDM scenario.}\label{f2}
\end{figure}
\begin{figure}
\centering
\includegraphics[scale=0.5]{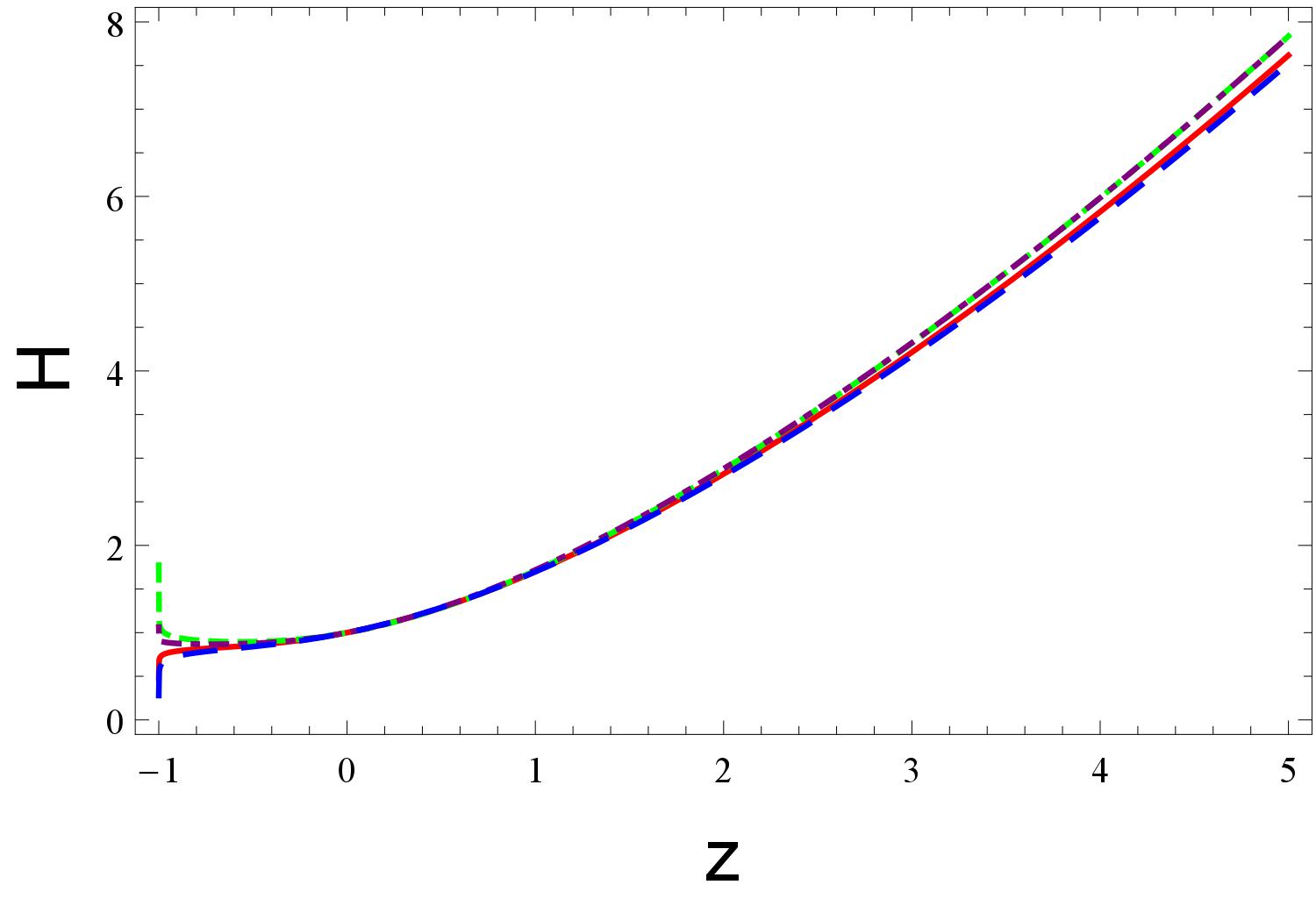}
\caption{The relation between the Hubble Parameter $H(z)$ and the redshift $z$ for the V-$\omega$CDM scenario, where the red (solid) line represents the background evolution constrained by the SNe Ia data-sets, the green (short-dashed) line SNe Ia+OHD, the purple (dash-dotted) line SNe Ia+OHD+WMAP-9+GW as well as the blue (long-dashed) line SNe Ia+OHD+Planck+GW.}\label{f3}
\end{figure}
\begin{figure}
\centering
\includegraphics[scale=0.5]{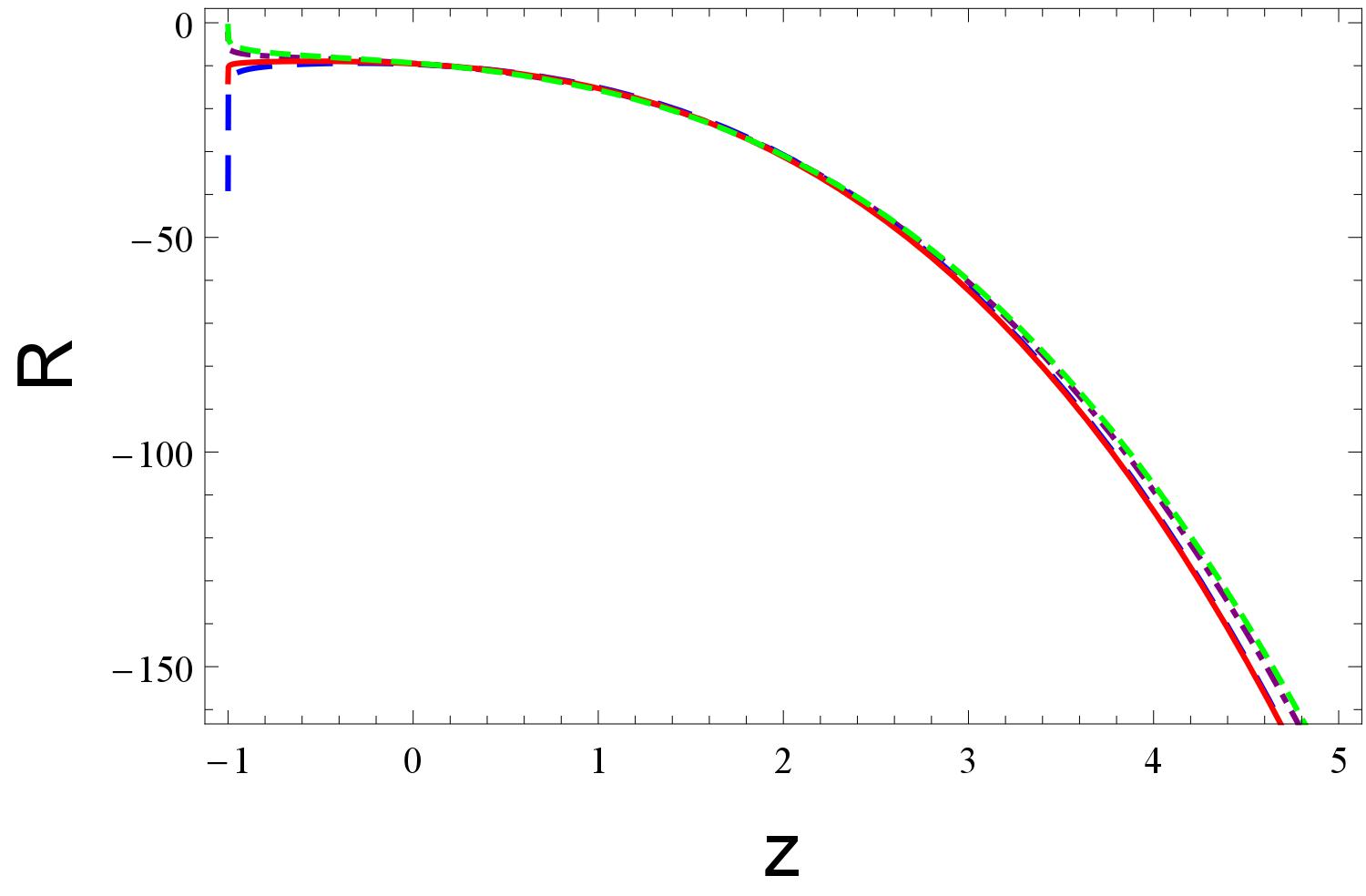}
\caption{The relation between the Ricci scalar curvature $R(z)$ and the redshift $z$, where the red (solid) line represents the reconstructed $f(R)$ theory  from the SNe Ia data-sets, the green (short-dashed) line from SNe Ia+OHD, the purple (dash-dotted) line from SNe Ia+OHD+WMAP-9+GW as well as the blue (long-dashed) line from SNe Ia+OHD+Planck+GW.}\label{f4}
\end{figure}
\begin{figure}
\centering
\includegraphics[scale=0.5]{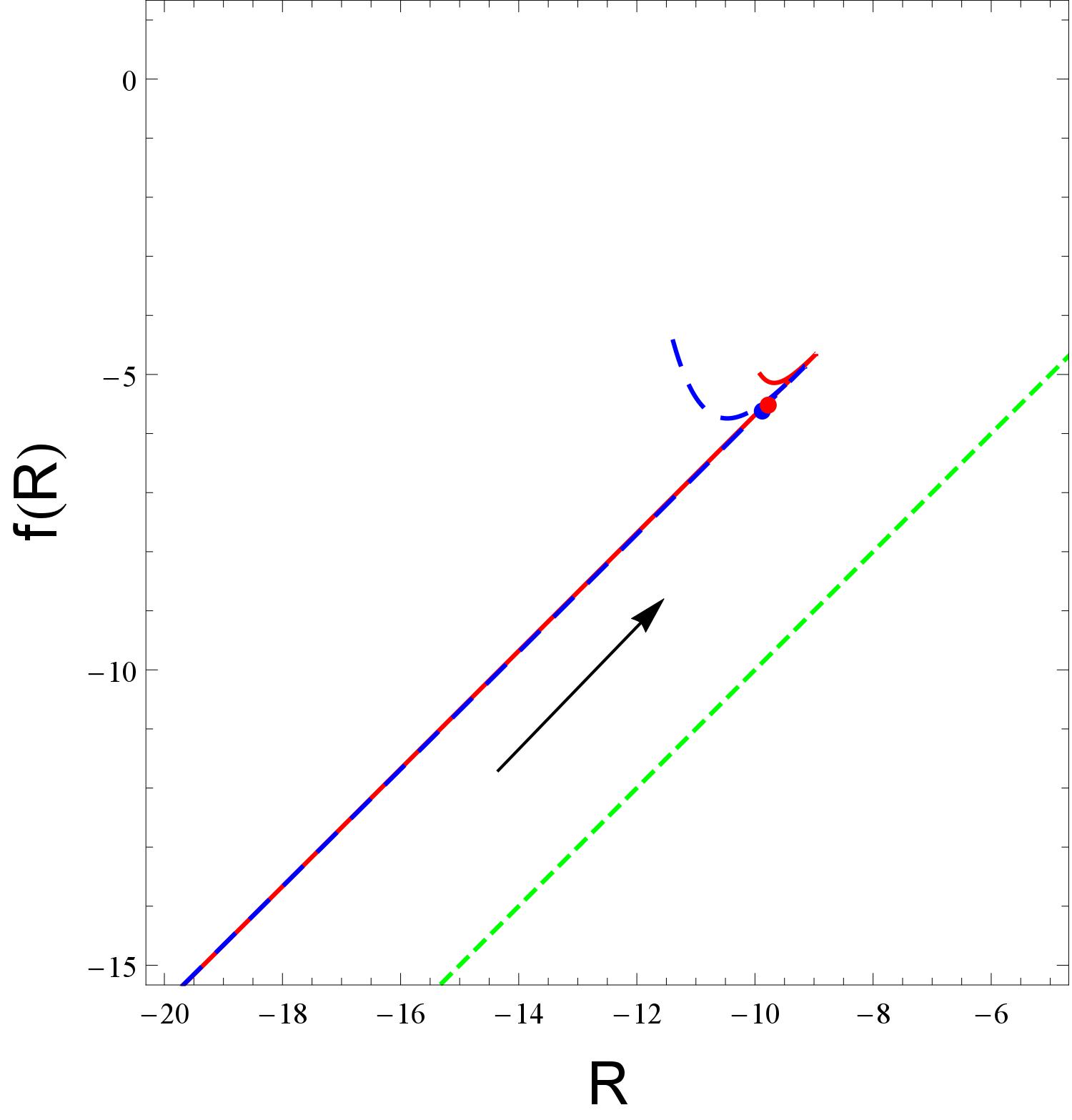}
\caption{The comparison between the reconstructed $f(R)$ theory with the V-$\omega$CDM scenario (solid line) and the reconstructed $f(R)$ theory with the $\omega$CDM scenario (dashed line) in the $f(R)-R$ plane, which only comes from the SNe Ia data-sets. The present epoch of reconstructed f(R) is shown as a dot.}\label{f5}
\end{figure}
\begin{figure}
\centering
\includegraphics[scale=0.5]{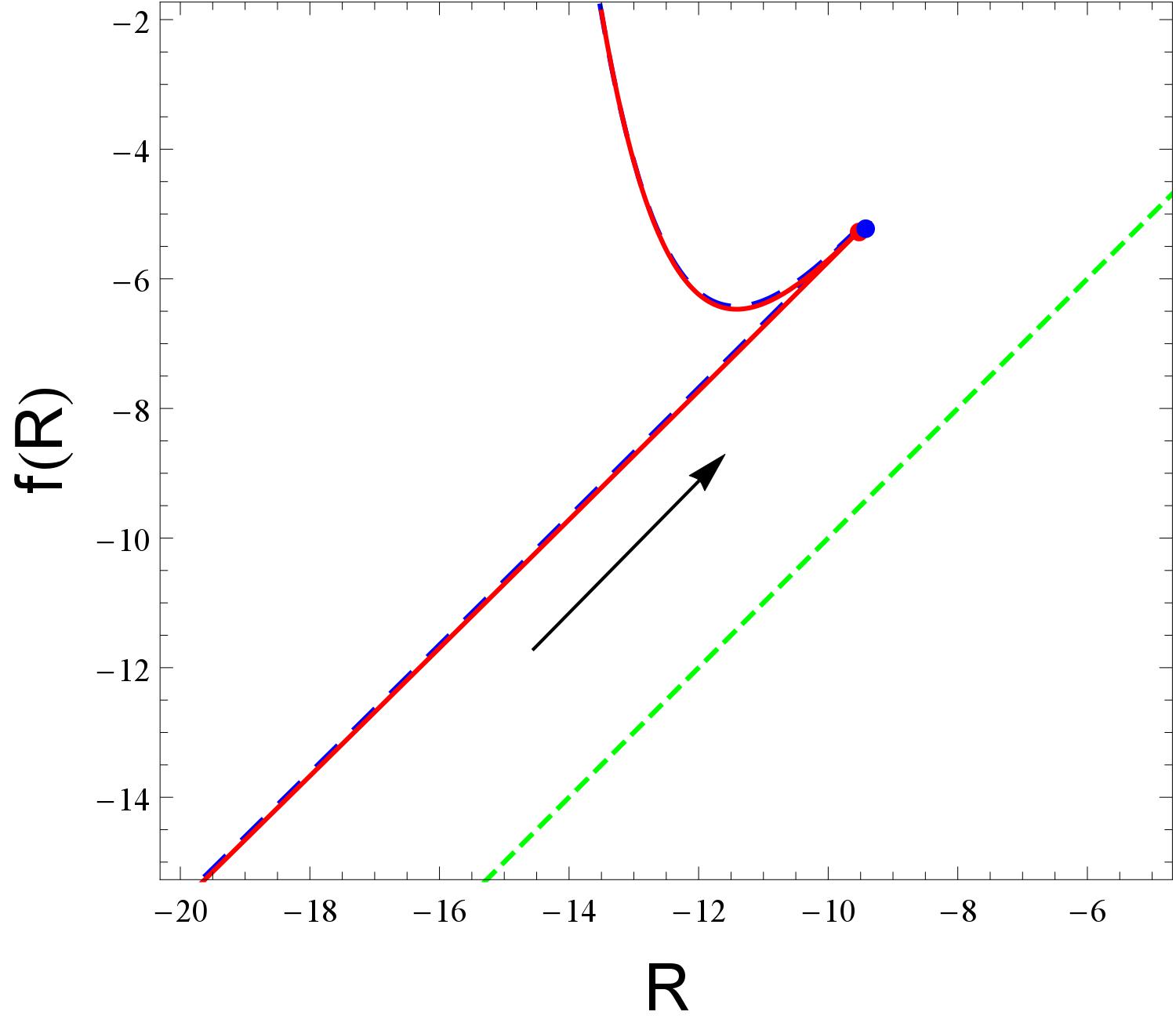}
\caption{The comparison between the reconstructed $f(R)$ theory with the V-$\omega$CDM scenario (solid line) and the reconstructed $f(R)$ theory with the $\omega$CDM scenario (dashed line) in the $f(R)-R$ plane, which comes from the SNe Ia+OHD+Planck+GW data-sets. The present epoch of reconstructed f(R) is shown as a dot.}\label{f6}
\end{figure}
Note that more details about calculations can be found in paper \cite{41}. More clearly, by solving the Eq. (\ref{28}) numerically, it is not difficult to determine what is the corresponding $f(R)$ theory of gravity for a given dark energy model. Before that, one must give out the reasonable boudary conditions with more physical considerations. According to the paper \cite{41}, three boudary conditions for Eq. (\ref{28}) can be expressed as :
\begin{equation}
f(z=0)=R_0+6H_0^2(1-\Omega_{m0}), \label{32}
\end{equation}
\begin{equation}
(\frac{df}{dz})_{z=0}=(\frac{dR}{dz})_{z=0}, \label{33}
\end{equation}
and
\begin{equation}
(\frac{d^2f}{dz^2})_{z=0}=(\frac{d^2R}{dz^2})_{z=0}. \label{34}
\end{equation}

Another important consideration is the dimensional analysis of different quantities in the adopted units ($8\pi G=c=\hbar=1$). From Eq. (\ref{20}), one can obtain the conclusion that the $f(R)$ has the same dimension as the Ricci scalar curvature $R$. Since $H$ and $R$ are described in the unit $s^{-1}$ and $s^{-2}$, one can easily derive the energy density like the critical density $\rho_{crit}=3H_0^2$ is in the unit $s^{-2}$. As a consequence, we measure time in the unit of $1/H_0$ in order that if $H_0=1$, all the quantities including $f(R)$ and $R$ will be dimensionless. Therefore, in this situation, we would like to utilize this useful property to solve Eq. (\ref{28}) numerically.

In Figure. \ref{f1}, the reconstructed $f(R)$ theories with the V-$\omega$CDM scenario from several different constraints are depicted. It is easy to be seen that there exists a substantially high degeneracy among these reconstructed $f(R)$ theories in the past, and around the present epoch,
they start exhibiting different behaviors. In the far future, one can also find that the reconstructed $f(R)$ theory using the SNe Ia data-sets and that using the SNe Ia+OHD+Planck+GW data-sets almost exhibit the same evolutional tendency, while the reconstructed $f(R)$ theory using the SNe Ia+OHD data-sets and that using the SNe Ia+OHD+WMAP-9+GW data-sets also exhibit the analogous behavior. In addition, the reconstructed $f(R)$ theories with the V-$\omega$CDM scenario are hardly distinguished at the present epoch. In Figure. \ref{f2}, we also depict the reconstructed $f(R)$ theories with the V-$\omega$CDM scenario in the $lf-lR$  plane ($lf=\log[-f(R)]$ and $lR=\log(-R)$) \cite{41} and obtain the completely same conclusion with that in Figure. \ref{f1}.

Since the aforementioned reconstructed $f(R)$ theories with the V-$\omega$CDM scenario from several different constraints perform very distinctively in the far future, it is worth investigating their background evolution behaviors. From Figures. \ref{f3} and \ref{f4}, it is very clear that the reconstructed $f(R)$ theory using the SNe Ia data-sets and that using the SNe Ia+OHD+Planck+GW data-sets have a very high degeneracy with each other during the evolutional process, in the meantime, the reconstructed $f(R)$ theory using the SNe Ia+OHD data-sets and that using the SNe Ia+OHD+WMAP-9+GW data-sets also exhibit the analogous evolutional tendency. Furthermore, one can explain the high degeneracy of the reconstructed $f(R)$ theories appearing in Figures. \ref{f1} and \ref{f2} very well by using this conclusion.

In Figure. \ref{f5}, we make a comparison between the reconstructed $f(R)$ theory with the V-$\omega$CDM scenario and that with the $\omega$CDM scenario constrained by the SNe Ia data-sets in the $f(R)-R$ plane. It is not difficult to discover that, the two different models can only be distinguished in the distant future and they are also hardly distinguished at the present stage, which means that the effect of bulk viscosity for $\omega$CDM scenario becomes obvious in the far future. Subsequently, we also make another comparison by using the SNe Ia+OHD+Planck+GW data-sets which has provided a substantially tight constraint for the two cosmological models, and find that the two models share a very high degeneracy, which implies the effect of bulk viscosity plays a negligible role for the V-$\omega$CDM model during the whole evolutional processes of the universe. Moreover, from Figures. \ref{f5} and \ref{f6}, one can also discover that the reconstructed $f(R)$ theories with the two models just deviate from the standard GR a little. Hence, the above-mentioned reconstructed $f(R)$ theories can act as reasonable $f(R)$ theories, which also indicates that the V-$\omega$CDM model and $\omega$CDM model can be regarded as reasonable dark energy models.

\section{Discussions and Conclusions}
The elegant discovery that the universe is undergoing a phase of accelerating expansion has inspired a great deal of researches to explore the cosmological origin and nature of the currently amazing phenomena. Due to a lack of deeper understanding at present, cosmologists have introduced an exotic energy component named dark energy to explain the cause of acceleration. Generally speaking, in the literature, there are two main approaches to understand the accelerated mechanism, i.e., the dynamical dark energy scenarios and the modified theories of gravity. Interestingly, there still exists a substantially degeneracy between the two approaches based on the currently astronomical observations. Therefore, we think it is very constructive and useful to investigate the relation between them.

In this paper, we mainly reconstruct the $f(R)$ theory of gravity from viscous cosmology. As a concrete instance, we explore the reconstruction of $f(R)$ using the  V-$\omega$CDM scenario. First of all, we constrain the V-$\omega$CDM scenario and $\omega$CDM scenario by utilizing several data-sets which includes the single data point from the newest event GW150914. We find that the joint constraints from SNe Ia+OHD+Planck+GW data-sets give out a tighter restriction for the parameters of the V-$\omega$CDM model than the left three joint constraints. It is noteworthy that the constraint result from SNe Ia+OHD+Planck+GW data-sets are stricter than that from SNe Ia+OHD+WMAP-9+GW data-sets may be attributed to the nice work found by Liu et al. \cite{53}. Subsequently, we carry out the reconstruction procedures by choosing the obtained best fitting values of the aforementioned two scenarios. It is easy to see that there exists a substantially high degeneracy among these reconstructed $f(R)$ theories in the past, and they start exhibiting different behaviors around the present epoch. In the distant future, one can also discover that the reconstructed $f(R)$ theory using the SNe Ia data-sets and that using the SNe Ia+OHD+Planck+GW data-sets almost exhibit the same evolutional tendency, while the reconstructed $f(R)$ theory using the SNe Ia+OHD data-sets and that using the SNe Ia+OHD+WMAP-9+GW data-sets also exhibit the analogous behavior. In addition, these reconstructed $f(R)$ theories with the V-$\omega$CDM scenario are hardly distinguished at the present epoch. In the $lf-lR$ plane, we also obtain the completely same conclusion as that in the $f(R)-R$ plane. Furthermore, through analyzing the background evolution of the reconstructed $f(R)$ theories, we can explain the high degeneracy reasonably which appears in Figures. \ref{f1}-\ref{f2}.

Another appealing and interesting point is to consider the role of the effect of bulk viscosity for the $\omega$CDM scenario during the cosmic evolution history. On the one hand, the constraint results indicate that the effect is very small for the V-$\omega$CDM scenario; on the other hand, by making a comparison between the reconstructed $f(R)$ theory with the V-$\omega$CDM scenario and that with the $\omega$CDM scenario constrained by using the SNe Ia+OHD+Planck+GW data-sets in the $f(R)-R$ plane, we find that the two scenarios share an extremely high degeneracy. Thus, we can obtain a conclusion that the effect of bulk viscosity plays a negligible role in the V-$\omega$CDM model during the whole evolutional processes of the universe. Note that in the present situation, we just take into account the case of a constant viscosity coefficient.

It is worth noticing that, in this case, we investigate the reconstruction of $f(R)$ theory by using the V-$\omega$CDM scenario and $\omega$CDM scenario in the standard framework of Einstein gravity, rather than other theories of gravity. Our goal is to explore and characterize the V-$\omega$CDM cosmological scenario in the framework of GR effectively in the point of view of modified gravities. As a concrete instance, we have studied the relationship between $f(R)$ gravity and V-$\omega$CDM scenario. We expect to constrain more dark energy models with higher precision observations in the future, in order that we can investigate the relationship between the dynamical dark energy models and the extended theories of gravity more accurately.

\section{acknowledgements}
During the preparation process of this paper, we are very grateful to Professors Saibal Ray, Bharat Ratra and Sergei. D. Odintsov for interesting communications on gravitational wave physics in cosmology and astrophysics. The author Deng Wang would like to thank Prof. Jing-Ling Chen for helpful discussions about quantum physics. This paper is partly supported by the National Science Foundation of China.

\end{document}